\documentclass[letterpaper,twocolumn,prb,floatfix,
  citeautoscript,showpacs,superscriptaddress]{revtex4-1}
\usepackage{epsfig,epstopdf}
\usepackage{graphicx}
\usepackage{amsfonts,amsmath,amssymb}
\usepackage{latexsym}
\usepackage{bm}
\usepackage{color}
\usepackage[
  colorlinks=true,
  allcolors=blue,
  breaklinks=true,
]{hyperref}

\begin{document}

\newcommand{\DyO}{Dy$_2$Ti$_2$O$_7$}
\newcommand{\HoO}{Ho$_2$Ti$_2$O$_7$}



\title{Monopole ordered phases in dipolar and nearest-neighbours Ising 
pyrochlore:\\ from spin ice to the ``all-in--all-out'' antiferromagnet}

\author{P. C. Guruciaga}
\email{pguruciaga@ifimar-conicet.gob.ar}
\affiliation{Instituto de Investigaciones F\'\i sicas de Mar del Plata
(IFIMAR), UNMdP-CONICET and\\
Departamento de F\'\i sica, Facultad de Ciencias Exactas y Naturales, 
Universidad Nacional de Mar del Plata, Dean Funes 3350, 7600 Mar del Plata,
Argentina.}

\author{S. A. Grigera}
\affiliation{Instituto de F\'{\i}sica de L\'{\i}quidos y Sistemas Biol\'ogicos
(IFLYSIB), UNLP-CONICET, 1900 La Plata, Argentina}
\affiliation{School of Physics and Astronomy, University of St Andrews, 
St Andrews KY16\ 9SS, United Kingdom}

\author{R. A. Borzi}
\affiliation{Instituto de Investigaciones Fisicoqu\'\i{}micas Te\'oricas y 
Aplicadas (INIFTA), UNLP-CONICET and\\
Departamento de F\'\i{}sica, Facultad de Ciencias Exactas, Universidad 
Nacional de La Plata,  c.c.\ 16, suc.\ 4, 1900 La Plata, Argentina}


\begin{abstract}
We study Ising pyrochlores by means of Monte Carlo simulations. We cover a set 
of exchange constants ranging from the frustrated ferromagnetic case (spin-ice)
to the fully-ordered ``all-in--all-out'' antiferromagnet in the dipolar
model, reinterpreting the results --as in an ionic 
system-- in terms of a {\it temperature vs. magnetic charge density} phase 
diagram. 
In spite of its spin nature and the presence of both double and single 
non-conserved magnetic charges, the dipolar model gives place to a phase diagram
which is quite comparable with those previously obtained for on-lattice systems
of electric charges, and on spin ice models with conserved number of single
magnetic charges. The contrast between these systems, to which we add results 
from the nearest-neighbours model, put forward other features
of our phase diagram --notably, a monopole fluid with charge order at high
monopole densities that persists up to arbitrarily high temperatures-- that can
only be explained taking into account construction constraints forced by the 
underlying spin degrees of freedom.    
\end{abstract}


\maketitle


\section{Introduction}    

Describing a complex material in terms of low-lying particle-like
collective excitations --\emph{quasi-particles}-- is one of the
key approaches in condensed matter physics \cite{anderson1997concepts}.
Phonons, magnons, electrons and holes
in semiconductors, are some of the better known examples of these
excitations. But quasi-particles form a very rich set, including
topological excitations and fractionalisation. In a first
approximation, these excitations are considered as
non-interacting, with all the complexity of the system hidden in
the quasi-particles themselves. The next layer of description,
where interactions between quasi-particles are included, leads to
a great variety of behaviours: from an-harmonic effects in crystals
to the stabilization of higher hierarchies of order such as magnon
binding \cite{khomskii2010basic} or magnon-mediated heavy fermion
superconductivity \cite{Steglich1979}. Recently, a new kind of
fractional point-like topological excitation has been proposed
theoretically \cite{Castelnovo2008} and evidence of its existence
found experimentally \cite{Morris2009,Kadowaki2009,Fennell2009} in the spin ice
compounds. These new quasi-particles are sources of magnetic field
and interact via a Coulomb-like potential, whence their name
magnetic monopoles \cite{Castelnovo2008}. This approach allows for a
very effective description of the thermodynamics, the dynamics and the out
of equilibrium behaviour of spin-ice systems
\cite{Castelnovo2008,Ryzhkin2005,Jaubert2009,Slobinsky2010,
Klemke2011,Petrenko2011,Jaubert2011,Castelnovo2011,Sen2014}.

\begin{figure}[htb]
\centering
\includegraphics[width=0.7\linewidth]{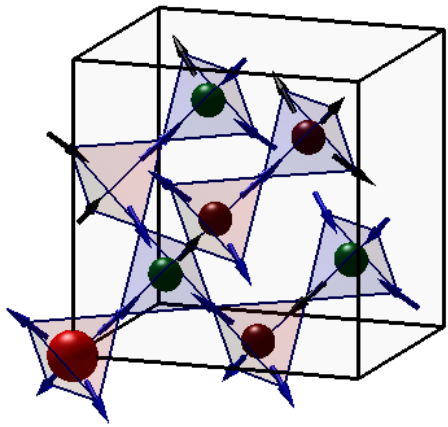}
\caption{Conventional unit cell of the pyrochlore lattice. 
In spin ice compounds, 
Ising-like spins occupy the vertices of corner-sharing ``up'' (pink)
and ``down'' (lilac) tetrahedra. Black/blue spins point inwards/outwards
(along the local $\langle 111 \rangle$ directions)
of an up tetrahedron. Via the dumbbell model \cite{Castelnovo2008},
the diverse spin configurations are mapped to
different types of magnetic charges: single/double (small/big) and 
positive/negative (green/red) monopoles
(note that since this unit cell is only a part of a bigger system, the 
constraint of local magnetic neutrality is not satisfied).
This configuration snapshot was obtained for a material with 
$J_{\textit{nn}}/D_{\textit{nn}} = -0.947$ at $T/ D_{\textit{nn}} \approx 0.57$,
resulting in a monopole density (defined as the average number of charges
per tetrahedron) of $\rho \approx 1.1$.}
\label{pyrochlore}
\end{figure}

The magnetic properties of spin ice materials can be
described by classical magnetic moments in a pyrochlore lattice,
occupying the vertices of corner-sharing tetrahedra. 
They behave at low temperatures as Ising-like spins \cite{Bramwell2001a} 
$\bm{\mu}_i=\mu S_i
\hat{\bf{e}}_i$ with $S_i=\pm 1$, pointing along the $\langle 111
\rangle$ directions $\hat{\bf{e}}_i$ (Fig.\ \ref{pyrochlore}). The magnetic
interactions of exchange and dipolar origin --of strengths $J$ and
$D$, respectively-- are well accounted for by the \emph{dipolar
spin ice model} (DSIM) Hamiltonian:

\begin{multline}
\frac{\mathcal{H}}{T}=\frac{D}{T} \biggl(\frac{J}{3D} \sum_{\langle ij\rangle}
S_i S_j + \\ + a^3 \sum_{(i,j)} \left[ \frac{\hat{\bf{e}}_i \cdot
\hat{\bf{e}}_j}{|{\bf{r}}_{ij}|^3}
           - \frac{3(\hat{\bf{e}}_i \cdot {\bf{r}}_{ij}) (\hat{\bf{e}}_j
      \cdot {\bf{r}}_{ij}) }{|{\bf{r}}_{ij}|^5} \right] S_i S_j \biggl)
\label{Hdip}
\end{multline}

\noindent where $T$ is the temperature, $a$ is the lattice spacing,
$\langle ij\rangle$ means that the sum is carried over nearest neighbours,
$|\mathbf r_{ij}|$ is the distance between spins $i$ and $j$ and 
$D=\mu_0\mu^2/(4\pi a^3)$.

The \emph{nearest neighbours spin ice model} (NNSIM) Hamiltonian is derived 
from the DSIM by keeping only the nearest neighbour contributions
of the dipolar interaction:

\begin{equation}
\frac{\mathcal{H}_{\textit{nn}}}{T}=\frac{J_{\textit{nn}}+D_{\textit{nn}}}{T}
\sum_{\langle ij\rangle} S_i S_j
\label{Hnn}
\end{equation}

\noindent with $J_{\textit{nn}}=J/3$ and $D_{\textit{nn}}=5D/3$, 
i.e.\ an effective exchange interaction
of strength $J_{\textit{eff}}=J_{\textit{nn}}+D_{\textit{nn}}$.
For the spin ice materials, $J_{\textit{eff}}$ in \eqref{Hnn} is
positive ($J_{\textit{nn}}/D_{\textit{nn}}>-1$) and imposes the
\emph{spin-ice rule} (named after Bernal and Fowler's ice rules
\cite{Bernal1933}): two spins should point in and two out of a
tetrahedron. This rule can be translated into field theory language as a
divergence free condition, which gives rise to a ``Coulomb phase''
\cite{Henley2010}.
Following Ref.\ \onlinecite{Castelnovo2008},
a violation of this law can be interpreted as the creation of a
charge --a monopole-- sitting in the tetrahedron; within the DSIM, a
Coulomb-like magnetic charge proportional to the divergence of the
spin vectors can be associated to each of these excitations. Single
excitations of opposite signs are related to ``3-in/1-out'' or
``1-in/3-out'' configurations, while double excitations (with double
charge) correspond to the ``all-in'' or ``all-out''
configurations. The antiferromagnetic version of this model, with
negative $J_{\textit{eff}}$ (i.e.\
$J_{\textit{nn}}/D_{\textit{nn}}<-1$), has an unfrustrated ground
state corresponding precisely to an ordered zinc-blende structure
of double charges: spins in alternating tetrahedra are in
configurations ``all-in'' and ``all-out''. Though much searched
for, no Ising pyrochlore with this spin ordering has been found
yet \cite{Matsuhira2009}.

At zero magnetic field the density of monopoles is regulated
by the sign and magnitude of $J_{\textit{eff}}/T$ which, in the
currently known materials, leads at best to moderately correlated
monopole fluids \cite{Zhou2011,Zhou2012}. In order to explicitly show the
effect of these correlations at low temperatures while stressing
the role of charge degrees of freedom, some of us recently
reported the results of simulations on a dipolar spin ice model
where a new ingredient was introduced: while keeping the dipolar
Hamiltonian \eqref{Hdip}, our approach in the \emph{conserved
monopole dipolar spin ice model} (CDSIM) was to use the density of
conserved single monopoles as the main control parameter \cite{Borzi2013}.
Our finding of phases with different degrees of
long range charge-like ordering reinforced the beauty and
simplicity of the monopolar scenario introduced by Castelnovo and
collaborators \cite{Castelnovo2008}. A very recent
contribution addresses this same issue in a wider scope 
\cite{Brooks-Bartlett2014}.
Excluding double charges in the
\emph{dumbbell model} (which --unlike the previous approach--
takes magnetic charges and not spins as their interacting simple entities),
they show that a Coulomb phase can still be defined
beneath a crystal of magnetic single charges. One drawback of
these approaches is that the explicit omission of \emph{double}
defects is somewhat contrived: in real materials, the limit
$J_{\textit{eff}}/T \to 0$ implies the proliferation of both
single and double excitations. Furthermore, within the CDSIM model,
the ordered ground state expected for negative
$J_{\textit{eff}}/T$ is not allowed \cite{DenHertogBC2000}.

In this paper we return to the usual DSIM in order to address
these shortcomings in the previous analysis and to extend it
in order to include the antiferromagnetic case.
Building up on previous results by den Hertog et al. \cite{DenHertogBC2000},
we show that when examined using the framework of monopolar excitations
both the DSIM and the CDSIM lead to the same physics, and in
particular to very similar phase diagrams.
Since within the NNSIM one would naively expect no monopole-monopole 
interaction, one would think charge degrees of freedom to play no role.
In spite of this, we will see that the phase diagram obtained for
the NNSIM model can be reinterpreted in terms of effective nearest-neighbours
interactions between double charges. We will also show that these effective 
interactions (neither dipolar nor exchange in origin, but arising from
correlations imposed by the internal degrees of freedom of the charges)
also affect the phase diagram in the presence of dipolar interactions.

\subsection*{Simulation details}    

We performed Monte Carlo simulations with single spin-flip
Metropolis algorithm, using Ewald summations to take into
account the long-range interactions 
\cite{Melko2004}.
We used a conventional cubic cell for the pyrochlore lattice,
which contains $16$ spins, and simulated systems with
$L\times L\times L$ cells. Thermodynamic data were
collected by starting at high temperatures and cooling very slowly,
for different values of $J_{\textit{nn}}/D_{\textit{nn}}$
(as this ratio will be negative throughout this work,
we will usually refer to its absolute value).
Typically, we needed $10^4$ Monte Carlo steps for equilibration and
$2\cdot 10^4$ for averaging at each temperature for $L=4$,
but we used up to $10^5$ steps for bigger lattices. First
order transition points deserve to be mentioned separately; there, full 
equilibration was only achieved after longer times (up to 
$5\cdot 10^5$ Monte Carlo steps) for sizes below $L = 5$.

\section{Charge degrees of freedom in the dipolar model}  

\begin{figure}[hbt]
\centering
\includegraphics[width=1.04\linewidth]{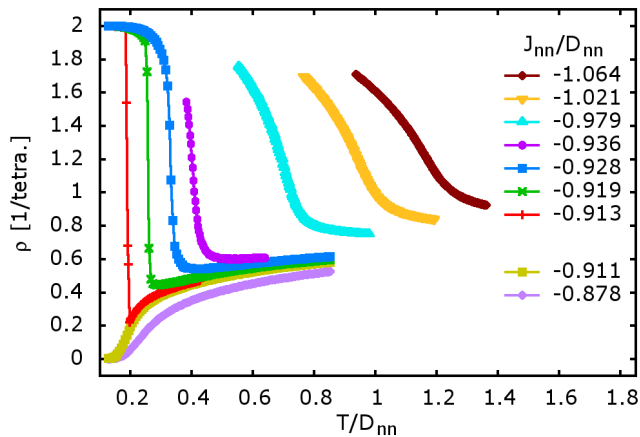}
\caption{Number of single charges per tetrahedron vs. 
temperature for $L=4$ and different values of
$J_{\textit{nn}}/D_{\textit{nn}}$. It is possible
to distinguish two very different regimes: some of the curves
(those with $|J_{\textit{nn}}/D_{\textit{nn}}|\le 0.911$) reach
$\rho=0$ when $T\to 0$, while the rest approach $\rho=2$.
Among the latter it is easy to notice that the jump between low
and high density becomes smoother as
$|J_{\textit{nn}}/D_{\textit{nn}}|$ increases.} \label{TvsRho_ew}
\end{figure}

Fig.\ \ref{TvsRho_ew} shows the monopole density $\rho$ as a
function of temperature for $L=4$. We define $\rho$ as the number
of single charges per unit tetrahedron, thus counting double
monopoles as the superposition of two single charges of the same
sign in a tetrahedron. In a similar way to Fig.\ 2 of Ref.\ 
\onlinecite{Brooks-Bartlett2014}, 
where only single charges were considered, we can distinguish two
families of curves according to their limit as $T\to 0$.
The first type, which tend to $\rho=0$ (with
$|J_{\textit{nn}}/D_{\textit{nn}}|\le 0.911$), were characterised
as spin ices by determining their residual entropy, via numerical
integration of the specific heat divided by temperature. The
second type are curves that tend to $\rho=2$ when $T\rightarrow
0$, meaning that they reach a state in which all the tetrahedra
are occupied by double monopoles (i.e.\ the non-frustrated
antiferromagnetic phase). For certain
values of $|J_{\textit{nn}}/D_{\textit{nn}}|$ in the latter group
of curves, the monopole density per tetrahedron suffers a sudden
change at low temperatures from $\rho\approx 0$ to $\rho\approx
2$. It is straightforward to
associate the appearance of this zinc-blende structure
\cite{Borzi2013} to the presence of Coulomb-like interactions between monopoles.
However, later we will show that this phase (which is no other than the 
``all-in--all-out'' phase mentioned
on the title and in the introduction) can also be
stabilized in the NNSIM, where these interactions are absent.

\begin{figure}[h!]
\centering
\includegraphics[width=\linewidth]{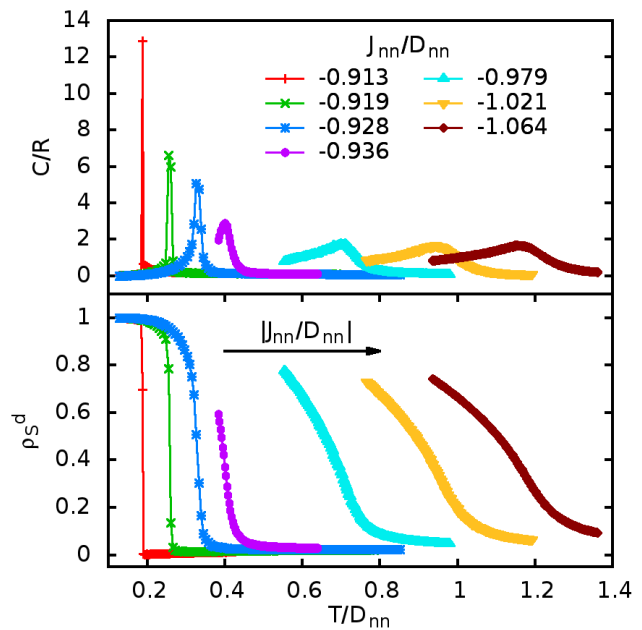}
\caption{Top: molar specific heat for $L=4$
as a function of temperature: the sharp, delta-like peak broadens
as $|J_{\textit{nn}}/D_{\textit{nn}}|$
increases --a sign of a first order phase transition that becomes second order.
This idea is reinforced by the double monopole staggered density (bottom),
which displays a step-like jump that becomes smooth and continuous.}
\label{todos_ew}
\end{figure}

The curves in Fig.\ \ref{todos_ew} (upper panel) represent the
molar specific heat $C$ as a function of temperature for different
values of $|J_{\textit{nn}}/D_{\textit{nn}}|\ge 0.913$ and $L=4$
(i.e.\ corresponding only to curves in the second branch of Fig.\ 
\ref{TvsRho_ew}).
We see sharp, delta-like peaks that become wider and shorter as
$|J_{\textit{nn}}/D_{\textit{nn}}|$ increases, correlated with the
jumps we noticed in $\rho$. With the development of a zinc-blende structure in
mind, we also present our results of the double monopole staggered
density $\rho_S^d$, defined as the average of the modulus
of the total magnetic charge due to double monopoles in up
tetrahedra per sublattice site per unit charge. A nearly zero
value of this quantity implies that there is no symmetry breaking
between the up and down tetrahedra sublattices, while
$\rho_S^d\approx 1$ is the result of a staggered ordering
of the magnetic charges: positive and negative double monopoles
alternating in the sublattices, occupying all the tetrahedra. In
Fig.\ \ref{todos_ew} (lower panel) we observe a jump between these
two states, which is sudden and step-like for low values of
$|J_{\textit{nn}}/D_{\textit{nn}}|$ and slowly changes into
continuous as that ratio increases. It is interesting to note that
these step-like jumps are correlated to the ones in $\rho$: the
system becomes dense and charge-ordered suddenly and
simultaneously, just as it happens in the crystallisation
transition of ionic fluids. On the other hand, the continuous
developing of both the staggered density and the number density at 
higher $|J_{\textit{nn}}/D_{\textit{nn}}|$ indicates the transition to a
phase in which the local density is homogeneous (fluid-like) but
charge-ordered.

\begin{figure}[hbt]
\centering
\includegraphics[width=\linewidth]{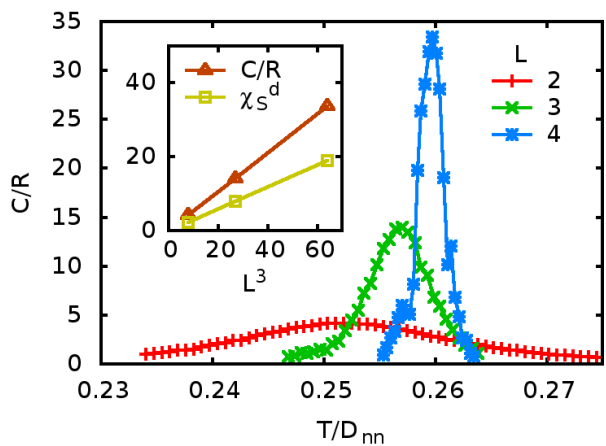}
\caption{For $J_{\textit{nn}}/D_{\textit{nn}} = -0.919$ 
the specific heat data
displays finite size effects consistent with a first order phase
transition. Moreover, the value of the specific heat and the double monopole
susceptibility at their maximum grow linearly with the volume of the
system, as shown in the inset.}
\label{tamfin1er}
\end{figure}

\begin{figure}[hbt]
\centering
\includegraphics[width=\linewidth]{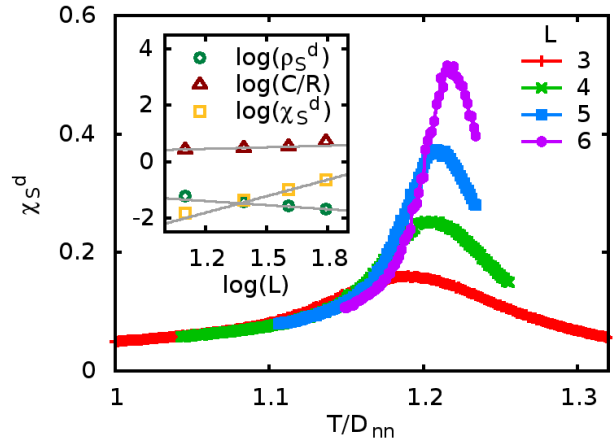}
\caption{For $J_{\textit{nn}}/D_{\textit{nn}}=-1.064$ the double
monopole susceptibility displays finite size effects consistent with a second
order phase transition. The values of $\rho_S^d (T_c(L))$,
$C (T_c(L))/R$ and $\chi_S^d (T_c(L))$ evolve as power laws with
the size of the system and present good correspondence with the 
behaviour expected for the three dimensional Ising model universality
class \cite{huang1987statistical} (grey lines).}
\label{tamfin2do}
\end{figure}

To confirm the existence of these transitions and determine their order
we performed finite size analysis over
values of $J_{\textit{nn}}/D_{\textit{nn}}$ representative of the two 
behaviours.
In Fig.\ \ref{tamfin1er} we present our results of the molar specific heat
for $J_{\textit{nn}}/D_{\textit{nn}} = -0.919$. The previous suggestion of a 
first order transition
taking place for these values of $J_{\textit{nn}}/D_{\textit{nn}}$
is backed up by the fact that the value of the specific heat
and the double monopole susceptibility
$\chi_S^d$ (defined as the fluctuations of the corresponding
staggered density over the temperature)
at their maximum are proportional to the volume of the
system (Fig.\ \ref{tamfin1er}, inset).
We also studied $J_{\textit{nn}}/D_{\textit{nn}}=-1.064$, in which the double 
monopole susceptibility
(Fig.\ \ref{tamfin2do}), as well as the specific heat and the double monopole
staggered density (not shown), evolve with the size of the system as in
a second order phase transition. The critical exponents are
consistent with the three-dimensional Ising universality class (Fig.\ 
\ref{tamfin2do}, inset).
Since this line of second order transitions becomes first order, there must be
a tri-critical point of the Blume-Emery-Griffiths universality class 
\cite{Blume1971}.

\begin{figure}[hbt]
\centering
\includegraphics[width=\linewidth]{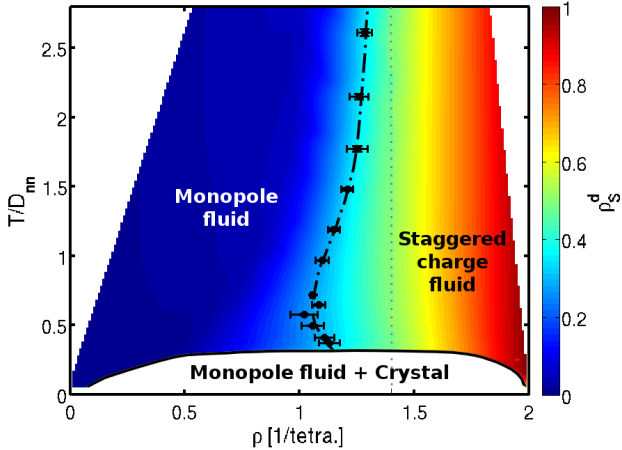}
\caption{Temperature vs. monopole density phase diagram with
dipolar (long-range) interactions. At low temperatures the system
undergoes a first order phase transition (white dome), where a low
density gas and a crystal coexist. The transition becomes second order
(dot-dashed line) at a tri-critical point near $\rho_t\approx 1.17$,
$T_t/D_{\textit{nn}}\approx 0.34$. 
The diagram is overlaid on top of an interpolated contour plot of
$\rho_S^d$ for $L=4$. The vertical dotted line at $\rho_c \approx 1.4$ 
indicates the critical density of the geometric transition in the NNSIM 
(see Sec.\ \ref{sec:GC}).}
\label{diag_rhos_ew}
\end{figure}

A nice way to summarise all the previous results
while emphasising the importance of charge degrees of freedom
is to construct a phase diagram in terms of the
monopole density and the temperature (Fig.\ \ref{diag_rhos_ew}).
The colour map corresponds to the interpolated value of the 
double monopole staggered density $\rho_S^d$ as a function of 
temperature and monopole number density (for $L=4$). 
The graph was obtained by combining a set of more than
thirty $\rho_S^d$ vs. $T$
curves (of which those shown on Fig.\ \ref{todos_ew}, bottom, 
constitute a subset) with $\rho$ vs. $T$ curves (like those on Fig.
\ref{TvsRho_ew}). The filled circles represent the location of the maximum 
on the specific heat for various values of 
$|J_{\textit{nn}}/D_{\textit{nn}}|$ (see Fig.\ \ref{todos_ew}, top).
Note that the vertical asymptote observed for this curve 
at high temperature corresponds to the limit in which antiferromagnetic 
exchange dominates over the dipolar interactions.
The white dome (drawn by estimating the region where
discontinuities would be observed for the infinite system) represents a
forbidden region in parameter space in which, in the thermodynamic
limit $L\to\infty$, the system cannot stay homogeneous. A system
with $\rho=0.4$ at $T/D_{\textit{nn}}\approx 0.22$ is exactly at the edge of the
dome; if cooled down, it would separate into two phases occupying different
parts of the material. On one hand, we would get a low-density monopole fluid, 
in which most of the tetrahedra are empty. On the other hand, a phase with
a high double monopole density would crystallise into the zinc-blende
ionic structure.
In a tri-critical point around $\rho_t\approx 1.17$ and
$T_t/D_{\textit{nn}}\approx 0.34$ the first order transition becomes second 
order, separating the monopole fluid and a staggered charge fluid
in which local density is homogeneous but negative and positive charges
display a tendency to occupy different sublattices,
thus breaking the symmetry without phase separation.
\footnote{It should be stressed that the right part of this diagram
corresponds to a single phase. The use of different names for two regions of
it --which follows Ref.\ \onlinecite{Borzi2013}-- does not imply that a 
different symmetry is broken, but intends to emphasise their different 
dynamical properties and the fact that, while the double monopole crystal 
may coexist with a low density charge-disordered fluid, the staggered charge 
fluid is homogeneous and occupies the whole space.}
It is interesting to note that both this staggered charge fluid and the crystal 
phase are more natural, double charge analogs of the single monopole ordered 
fluid and crystal found in the slightly artificial CDSIM (see Fig.\ 4 of Ref.\ 
\onlinecite{Borzi2013}). The apparent differences between this phase diagram 
and that presented on Fig.\ 1 of Ref.\ \onlinecite{DenHertogBC2000} 
(of which we give our own version on Fig.\ \ref{diag_TvsJ}) should not mask the 
fact that they both describe the same physics.

On the other hand, despite the great similarity between the phase diagram of Fig.\ 
\ref{diag_rhos_ew}, and that for the CDSIM, several differences arise. 
Since we used a single 
spin flip algorithm and did not fix the number of monopolar defects,
the Melko-Gingras-den Hertog
first order transition \cite{Melko2004} at $T/D_{\textit{nn}}=0.077$ to an
``ordered vacuum'' of magnetic charges \cite{Borzi2013} is not visible.
Also, Fig.\ \ref{diag_rhos_ew} exhibits a re-entrant behaviour of the
monopole fluid at temperatures just above $T_t/D_{\textit{nn}}$.
Finally, one would expect that a disordered phase should always
be found at high temperatures ($T/D_{\textit{nn}} \gg 1$)
for any charge density, when entropic forces overcome monopole attraction.
While this is true both for a system of real Coulomb charges in a lattice
\cite{Dickman1999} and for the CDSIM \cite{Borzi2013},
we do not see the line of second order transitions
joining $\rho = 2$ in the present case.
Instead, this line reaches an asymptotically vertical behaviour,
parallel to the temperature axis. This fact, which seems to point to an
infinitely high interaction energy between charges, could hardly be explained
by the monopolar Hamiltonian on its own \cite{Castelnovo2008}. These last two 
cases deserve special attention, and will be analysed in depth in the next 
sections.

\section{Comparison between the dipolar and the nearest-neighbours model}   

In order to better understand the limits of the magnetic charges picture
and the energetics of the models we have introduced,
we also studied the NNSIM (Eq.\ \eqref{Hnn}) for
different ratios $J_{\textit{nn}}/D_{\textit{nn}}$. Similar to the
previous case, we found a
peak in the specific heat and the double
monopole susceptibility, and a steep rise in the double
monopole staggered density as a function of
temperature (not shown), suggesting a phase transition.
In this case the behaviour of these quantities is typical
of a second order transition
for each value of $J_{\textit{nn}}/D_{\textit{nn}}$ studied,
with no delta-like peaks or discontinuities.

\begin{figure}[hbt]
\centering
\includegraphics[width=\linewidth]{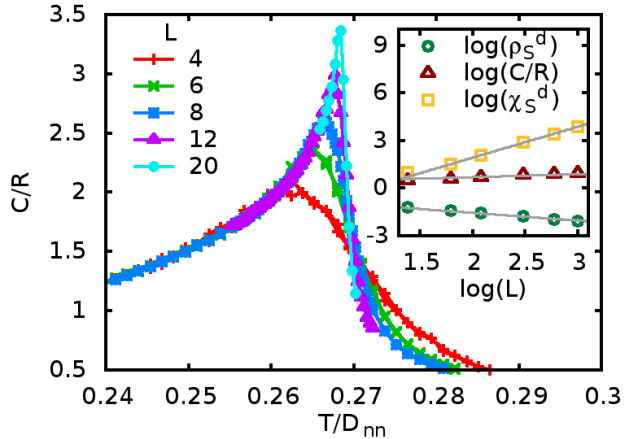}
\caption{In the case of the nearest neighbour model, the 
specific heat
displays finite size effects characteristic of a second order phase transition
for $J_{\textit{nn}}/D_{\textit{nn}}=-1.064$. Inset: again, the values of 
$\rho_S^d (T_c(L))$,
$C (T_c(L))/R$ and $\chi_S^d (T_c(L))$ evolve as power laws with
the size of the system and present good correspondence with the
behaviour expected for the three dimensional Ising model 
\cite{huang1987statistical} (grey lines).}
\label{nn_tamfin_2do}
\end{figure}

\begin{figure}[hbt]
\centering
\includegraphics[width=\linewidth]{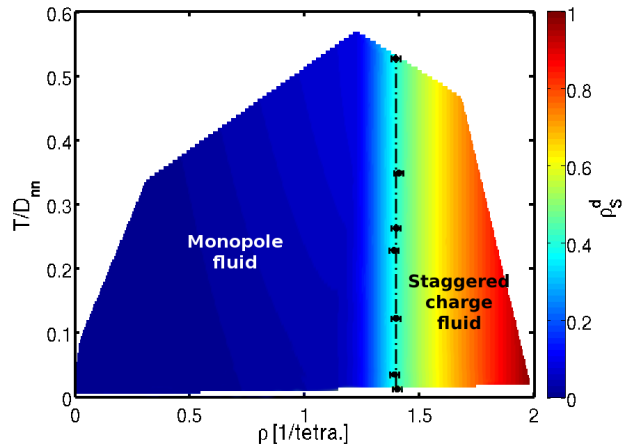}
\caption{Temperature vs. monopole density phase diagram with 
nearest-neighbours interactions. The second order phase transition 
(dot-dashed line) between the 
gas and the staggered charge fluid occurs at $\rho_c\approx 1.4$ at all 
temperatures, hence we call it a geometric phase transition. The diagram is 
overlaid on top of an interpolated contour plot of $\rho_S^d$ for 
$L=4$.}
\label{diag_rhos_nn}
\end{figure}

In Fig.\ \ref{nn_tamfin_2do} we show the peak in the the molar specific heat 
for $J_{\textit{nn}}/D_{\textit{nn}}=-1.064$ and different sizes of the system.
The inset shows that its value at the critical temperature
--as well as the order parameter $\rho_S^d$ and its
fluctuations $\chi_S^d$-- evolves with size as a power law.
As expected, we determined that the transition again belongs to the 
three-dimensional Ising universality class. Furthermore, since
only one energy scale is present in this model (see Eq.\ \eqref{Hnn}), the $T$ 
vs. $\rho$ phase diagram shown in Fig.\ \ref{diag_rhos_nn} does not depend on 
$T/D_{\textit{nn}}$. 

\begin{figure}[hbt]
\centering
\includegraphics[width=0.95\linewidth]{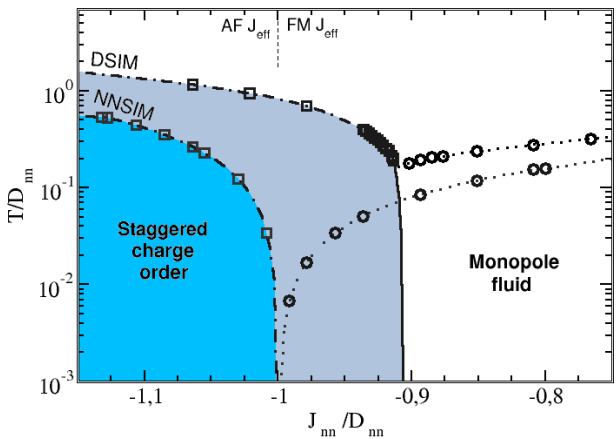}
\caption{
$T/D_{\textit{nn}}$ vs. $J_{\textit{nn}}/D_{\textit{nn}}$ 
phase diagram for both dipolar and
nearest-neighbours spin ice models. Dotted lines represent crossovers,
while solid and dot-dashed lines stand for first and second order phase 
transitions, respectively. The effect of the dipolar interactions is clearly 
seen: the system orders at the alternating charge configuration even for values
of $J_{\textit{eff}}>0$ because the Coulomb attraction between monopoles of 
opposite charge makes it energetically favourable to create and order these 
particles.
Melko-Gingras-den Hertog first order transition \cite{Melko2004} is not visible
because we used a single spin-flip algorithm
that mimics the real material and thus freezes at low temperatures.}
\label{diag_TvsJ}
\end{figure}

The dependence of the critical temperature on the
relative strength of the exchange and dipolar interactions 
has already been explored for the DSIM \cite{DenHertogBC2000,Melko2004}.
In order to explicitly show the effect
of dipolar interactions, we now present our version  of the 
$T/D_{\textit{nn}}$ vs. $J_{\textit{nn}}/D_{\textit{nn}}$ phase diagram 
for both the DSIM and the NNSIM
(Fig.\ \ref{diag_TvsJ}). The first thing we note is that in the DSIM the
staggered charge fluid
extends to a region with small but ferromagnetic $J_{\textit{eff}}$
(or, equivalently, $J_{\textit{nn}}/D_{\textit{nn}}\agt -1$); this does
not happen in the NNSIM. 
This can be easily understood within the monopole picture:
in the same way as neutral atoms of {\it Na} and {\it Cl} ionize
to form {\it NaCl}, the energy spent in the creation of the
monopoles can be compensated by the attraction between them.
As a matter of fact, comparing the Madelung energy of 
a zinc-blende lattice of
oppositely charged double monopoles and the energy needed to create these 
charges out of the vacuum \cite{Castelnovo2008} 
results in a zero-temperature limiting value of
$J_{\textit{nn}}/D_{\textit{nn}} \approx -0.918$ \cite{Brooks-Bartlett2014},
which is reasonably close --within the monopole picture approximation--
to the value $-0.905$ found by us (Fig.\ \ref{diag_TvsJ})
and Ref.\ \onlinecite{Melko2004} --both in spin systems.
This quantitative agreement provides additional support for the 
monopole picture of spin ice.

\section{Role of the spin degrees of freedom}  

\subsection{Charge interactions vs. correlations}  
\label{sec:GC}

The strength of the monopole picture, that allows not only a qualitative
but also a quantitative understanding of the phase diagram, seems at first 
sight to be weakened by a plain fact: no Coulomb attraction
between oppositely charged monopoles is expected in the NNSIM 
(note that we continue calling the excitations ``monopoles'' despite
the fact that they have lost their main quality as charges). 
While it is simple to understand why the spins would order in the alternating 
``all-in--all-out'' configuration when only a negative nearest-neighbours 
interaction $J_{\textit{eff}}$
is considered (Fig.\ \ref{diag_TvsJ}), it is much harder
to rationalise, limiting ourselves to the monopole picture, why non-attracting 
monopoles would experience any sort of staggered ordering.

This simple puzzle can be solved by taking into account the \emph{correlations} 
between monopoles, which transcend any Coulomb-like interactions.
It is not energetic disfavour, but construction constraints that prevent more 
than one double charge in a single diamond site.
In the same way, the underlying spin configuration (``all-in'' or ``all-out'') 
forbids two double monopoles with the
same double charge to be placed in adjacent tetrahedra.
A similar reason makes more probable to find single monopoles with the opposite 
charge around any double monopole. Provided there are enough double charges in a
diamond lattice, this correlation --quite equivalent to an infinite
repulsion between double charges of the same sign at nearest neighbour sites-- 
will induce staggered ordering on
the system, \emph{irrespective of the system temperature}.

\begin{figure}[hbt]
\centering
\includegraphics[width=\linewidth]{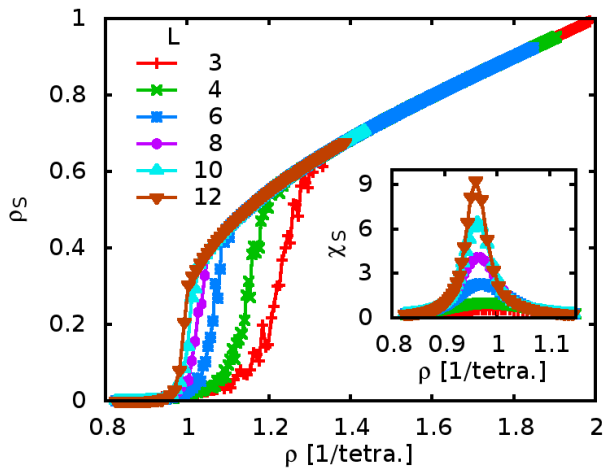}
\caption{Staggered charge density (main figure) and 
its fluctuations (inset) vs. 
particle density for a system of ``plus'' and ``minus'' non-interacting hard
spheres within the Grand Canonical Ensemble. 
Finite size scaling of these quantities is consistent with the three dimensional
Ising universality class \cite{huang1987statistical} (not shown).}
\label{GC}
\end{figure}

In order to test this point quantitatively, we have simulated a system of
nominal ``plus'' and ``minus'' hard spheres
in the diamond lattice within the Grand Canonical Ensemble.
In analogy to the NNSIM, we included no
interactions between spheres, but a
constraint was imposed forbbiding two like spheres
to be placed in neighbouring sites. Figure \ref{GC}
shows the staggered charge density $\rho_S$ for the spheres as a 
function of total average sphere density
\footnote{In order to make a closer analogy between the hard spheres and the 
double charges, we have given each sphere a value of $2$ in number, thus 
reaching a saturation value for the density of $\rho=2$.} $\rho$, 
for different system
sizes --a cubic unit cell of side $L$ is again implied. The sudden increase
in $\rho_S$ reflects the charge-like order being stabilised in the 
system as a consequence of sphere correlations.
A previously known example of this kind of symmetry breaking in
hard-sphere systems can be found in binary mixtures with a radius ratio
of $R_A/R_B\approx 0.4$ and $0.76$, where $R_A$ ($R_B$) is the radius of 
the small (large) spheres. Driven by differential excluded volume effects,
these systems present $NaCl$ and $CsCl$ structures, respectively 
\cite{Kummerfeld2008,Filion2009}, in which 
every $A$ particle is surrounded only by $B$ particles, and vice versa,
resembling the staggered ordering of our non-interacting double 
``monopoles''.

Once more, finite size scaling of the order parameter and its fluctuations
(Fig.\ \ref{GC}, inset) allowed us to identify the three dimensional Ising 
universality class.
The extrapolated critical density $\rho_c(L \to \infty) = 0.964 \pm 0.004$
is near the critical value found for $\rho_c$
in Fig.\ \ref{diag_rhos_nn} for the NNSIM.
The differences between these two critical monopole concentrations
can be explained realising that our simple sphere model
does not consider single monopoles, which add
extra charge with smaller correlations.
Considering then that even the NNSIM involves
an effective contact interaction between monopoles,
we can now understand the independence of $\rho_c$
with temperature observed in Fig.\ \ref{diag_rhos_nn}
as a result of a construction constraint.

This reasoning gains in depth when we contemplate its implications to the 
dipolar model studied before.
While the low temperature region of the phase diagram in Fig.\ 
\ref{diag_rhos_ew} is, as discussed,
obviously dominated by the Coulomb attraction between monopoles, the ordered
phase seen for $T/D_{\textit{nn}} \gg 1$ at high monopole density can now be 
trivially explained in terms of the aforementioned charge correlations. Indeed,
the vertical asymptote we noticed in Fig.\ \ref{diag_rhos_ew} shows the
equivalence between the DSIM and NNSIM in the
limit of high $T/D_{\textit{nn}}$. 

As mentioned, correlations beyond charge interactions are much smaller for 
single charges.
Indeed, using our conserved-monopole algorithm \cite{Borzi2013} for nearest 
neighbours spin interactions we have proved that a system of single monopoles 
remains charge-disordered, even in the limit of $\rho=1$. This explains the 
observation of a high temperature limit for the stability of the fluid of 
single monopoles observed in Ref.\ \onlinecite{Borzi2013}.

\subsection{Discussion: re-entrance near the tri-critical point} 

Figure \ref{diag_rhos_ew} suggest the re-entrance of the charge-disordered phase
near the dome.
Indeed, we have observed that this behaviour persists for bigger lattices, of 
which we have explored up to $L=5$. Though not in a conclusive way, this 
suggests that this behaviour is not the mere consequence of the tri-critical 
fluctuations. As in the previous discussion, we propose an explanation for the 
re-entrant behaviour --which is not observed in the Blume-Emery-Griffiths model 
\cite{Blume1971} nor in lattice models of real charges
\cite{Dickman1999,Kobelev2002} nor in the CDSIM \cite{Borzi2013}-- 
which transcends the mere charge-like degrees of freedom. 

A clue of the entropic origin of the re-entrance can be obtained
by noticing that the charge-ordered phase extends its region of stability
on increasing temperature, reaching lower densities. We have checked that
along the transition line the reduction in $\rho$ is related with an increase 
in the number of
empty (``2-in/2-out'') tetrahedra, while keeping an approximately constant
concentration of single-charge monopoles. This
mechanism, which necessarily decreases the number of double-defects, 
is very efficient in increasing the entropy of the system. This entropic 
contribution --which can be related to \emph{internal} degrees of freedom of 
the charges-- can outweigh the expected decrease in energy, reducing the free
energy of the staggered phase. The balance is no longer
possible for low enough densities, and the previously 
observed charge-like behaviour \cite{Dickman1999,Kobelev2002,Borzi2013}
is recovered, with $\rho_c$ increasing with temperature.

\subsection{Charge order with spin disorder: best chance within an Ising 
pyrochlore material} 

We have already mentioned the impossibility of 
stabilising in a real Ising pyrochlore material a perfect single-monopole 
crystal at zero magnetic field. 
As stated in Ref.\ \onlinecite{Brooks-Bartlett2014}, such a crystal would imply 
the existence of charge order coexisting with a Coulomb phase. We believe that 
the staggered charge-ordered phase near the re-entrant region is the closest we 
can get in this system to this situation. 
At $T/D_{\textit{nn}} \approx 0.57$ and $\rho \approx 1.1$ 
(see Fig.\ \ref{pyrochlore} for a snapshot of a part of the system at these 
values) almost $50\%$ of
the diamond lattice sites are occupied by single-defects, while only $30 \%$
are double monopoles.
This leftmost transition point, in which the staggered charge-order
is impending, is quantitatively compatible with the phase diagram obtained for
the CDSIM (since only single charges are allowed in this model, a double 
monopole in the DSIM must be equated to a single monopole in the CDSIM, then the
total density $\rho = 1.1$ in the first model corresponds to 
$\rho \approx 0.5+0.3 = 0.8$ in the latter).
Indeed, Fig.\ 4 in Ref.\ \onlinecite{Borzi2013} shows that for $80 \%$ 
occupation staggered order is established near 
$T/D=5/3\text{ }T/D_{\textit{nn}}\approx 0.95$, plainly consistent with the 
value  $T/D_{\textit{nn}}\approx 0.57$
we find for the unrestricted DSIM. This coincidence implies that double 
monopoles for this level of dilution
contribute to the staggered-charge order similarly to single monopoles.
An important consequence of this fact is that a neutron scattering measurement 
of an Ising pyrochlore material with 
$J_{\textit{nn}}/D_{\textit{nn}} \approx -0.947$ at 
$T \approx 0.57 D_{\textit{nn}}$ (so as
to have $\rho \approx 1.1$ inside the ordered phase, but with a significant 
fraction of single-charged monopoles) should give a structure factor quite 
similar to that simulated in Fig.\ 4 of Ref.\ \onlinecite{Brooks-Bartlett2014}. 
This pattern
shows signs of spin fragmentation, combining Bragg peaks
from an ``all-in--all-out'' structure with the pinch points
which characterise the underlying Coulomb phase.

\section{Conclusions} 

The path we have taken in this work has been two-fold. Firstly, we have shown 
how the monopole picture arises naturally as a tool to reinterpret and 
understand the physics of dipolar Ising pyrochlore systems, encompassing both 
spin ice and antiferromagnetic materials. The magnetic phase diagram for 
classical Ising pyrochlores calculated by den Hertog and collaborators 
\cite{DenHertogBC2000}, separating the ``all-in--all-out'' antiferromagnet from 
the spin ice systems in terms of the effective exchange constant values, has 
been recast into {\it temperature vs. density}, in analogy to the phase 
diagrams in molecular systems. Notably, in spite of its underlying spin nature 
and the presence of four types of different non-conserved magnetic charges 
(plus and minus, single and double), the dipolar model gives rise to a  phase 
diagram which is quite comparable with those previously obtained for on-lattice 
systems of electric charges \cite{Dickman1999}, and on spin ice models with 
conserved number of single magnetic charges \cite{Borzi2013}. Thinking just on 
monopoles and their interactions made it simple to justify quantitatively the 
extent to which the antiferromagnetic phase gets into the region of 
ferromagnetic first neighbours effective exchange interaction, by evaluating the
Madelung energy of a double monopole crystal. This plainly exemplifies how
the power of the picture does not limit itself to the qualitative understanding 
of the phases present. Within the nearest-neighbours model, the stabilisation of
the antiferromagnetic phase was interpreted within the monopole picture in terms
of the proliferation of monopoles with no Coulomb interactions but which 
implicitly force a nearest-neighbour exclusion condition between like types. 
This exclusion condition mimics an attraction/repulsion between like/different 
charges, maintaining the idea of monopole as a useful concept even within the 
NNSIM.

We also encountered certain aspects on the phase diagrams which cannot be 
explained in terms of simple charges. Like the discovery of internal degrees of 
freedom in particles previously thought as indivisible building blocks, these 
findings are far from making the monopole picture less interesting. 
Among these peculiarities we found a re-entrance of the disordered fluid, and a
staggered charge-ordered fluid phase which --for high enough densities-- can be 
stable at temperatures arbitrarily much higher than the energy scale 
characterising charge interactions. This second fact could be understood in 
terms of construction constraints --the exclusion condition we referred to in
the previous paragraph-- inherent to the spin nature of the excitations. 
The local constraint (as opposed to Coulomb-like monopole attraction) is the 
predominant correlation mechanism between monopoles at high temperatures and 
high monopole density, and the only interaction for the NNSIM. 
We explained the re-entrance noticing that lowering the density of double 
charges induces a reduction in the energy, but a boost in this energy's
degeneracy. A final remark is the identification near the re-entrance of a 
good candidate for a state where charge order can coexist with a Coulomb phase 
\cite{Brooks-Bartlett2014} (an Ising pyrochlore material with 
$J_{\textit{nn}}/D_{\textit{nn}} \approx -0.947$, at $T/D_{\textit{nn}} 
\approx 0.57$). 
The scattering pattern of this state should show signs of spin fragmentation, 
combining Bragg peaks from an ``all-in--all-out'' structure with the pinch 
points which characterise the underlying Coulomb phase 
\cite{Brooks-Bartlett2014}.

\begin{acknowledgments}
We thank T.\ S.\ Grigera for helpful discussions. 
This work was supported by Consejo Nacional de Investigaciones Cient\'\i ficas
y T\'ecnicas (CONICET) and Agencia Nacional de Promoci\'on Cient\'\i fica
y Tecnol\'ogica (ANPCyT), Argentina.
\end{acknowledgments}



%

\end{document}